\documentclass[11pt]{article}
\usepackage{moriond,epsfig}

\def\Journal#1#2#3#4{{#1} {\bf #2}, #3 (#4)}

\def\AP{\em Ann. Phys.}
\def\JHEP{\em JHEP}

\def\PRL{\em Phys. Rev. Lett.}
\def\PRD{{\em Phys. Rev.} D}
\def\RMP{\em Rev. Mod. Phys.}
\def\RPP{\em Rept. Prog. Phys.}

\def\be{\begin{equation}}
\def\ee{\end{equation}}
\def\bea{\begin{eqnarray}}
\def\eea{\end{eqnarray}}

\begin{document}
\vspace*{4cm}
\title{SEARCH FOR WMAP-COMPATIBLE SIMPLE $SO(10)$ SUSY GUTs}

\author{ SEZEN SEKMEN }

\address{Department of Physics, Middle East Technical University,\\
TR-06531, Ankara, Turkey}

\maketitle\abstracts{
Unification of GUT-scale $t-b-\tau$ Yukawa couplings is a significant feature of simple $SO(10)$ SUSY 
GUTs. Here we present the results of a search that used the Markov Chain Monte Carlo technique to 
investigate regions of Yukawa unification and WMAP-compatible dark matter relic density in $SO(10)$-like 
MSSM parameter spaces.  We mention the possible LHC signatures of Yukawa unified scenarios and discuss 
the consequences for dark matter.} 

\section{Introduction}

Grand Unification is regarded as an inspirational ingredient of models that claim to explain the 
fundamental laws of nature.  A highly motivated scenario in this context originates from grand 
unification via the $SO(10)$ gauge group~\cite{so10}.  Simple supersymmetric implementations of $SO(10)$ 
GUTs unify all matter fields in each generation within a 16-dimensional irreducible representation and 
two Higgs doublets of the MSSM within a 10-dimensional irreducible representation.  Such a formalism 
automatically includes heavy right-handed neutrino states and the resulting structure of the neutrino 
sector implies a successful theory of baryogenesis via intermediate scale leptogenesis.  Moreover the 
$SO(10)$ models are left-right symmetric and this enables them to provide a solution to the strong CP 
problem and to naturally induce R-parity conservation.

Besides gauge coupling unification, $SO(10)$ SUSY GUTs additionally require the unification of 
3rd generation Yukawa couplings at the GUT scale $(M_{GUT})$.  This is explicitly seen from the 
expression of the superpotential above $M_{GUT}$, which takes the form $\hat{f} \ni f \hat{\psi}_{16} 
\hat{\psi}_{16} \hat{\phi}_{10} + \cdots$.  An exact unification occurs at tree level while several  
percent corrections arise at the loop level.  As a result we can assume that any sign of Yukawa 
unification from observations could be a hint to the existence of $SO(10)$ SUSY GUTs. 

Our aim in the study is to investigate the characteristics that arise in a SUSY model when GUT scale 
Yukawa unification is imposed and to determine the experimental signatures that would distinguish such 
Yukawa-unified models from the others.  In this context we assume a theoretical framework where nature 
is explained by an $SO(10)$ symmetry above $M_{GUT}$.  Then at $M_{GUT}$, $SO(10)$ breaks to MSSM plus 
some heavy right-handed neutrino states.  At the weak scale, the content of the theory is equivalent to 
that of the MSSM.  

The GUT scale soft SUSY breaking parameters are constrained by the requirement of the $SO(10)$ symmetry.  
Unified representations would favor common SSB masses $"m_{16}"$ for the scalars and $"m_{10}"$ for the 
Higgses, but in order to achieve REWSB, SSB Higgs masses should be split, satisyfing $m_{H_d} > 
m_{H_u}$.  Here we examine two different methods to generate the necessary Higgs splitting: The first 
approach defines the Higgs masses as $m^2_{H_{u,d}} = m^2_{10} \mp 2M^2_D$.  Here splitting is 
parametrized by $M_D$, which is the magnitude of the D-terms in the scalar potential of the extra $U(1]$ 
group that is a by-product of the $SO(10)$ breaking.  The parameters of this GUT scale Higgs input (GSH) 
scenario are
\begin{equation}
m_{16},\,m_{10},\,M_D^2,\,m_{1/2},\,A_0,\,\tan\beta,\,sgn(\mu)
\end{equation}
The socond approach was put forward in order to generate Yukawa unified solutions with low $\mu$ 
parameter and low $m_A$.  Such solutions were found to exist by Blaszek, Dermisek and Raby at a study 
where they assumed perfect Yukawa unification at $M_{GUT}$ and made a fit to the weak scale 
observables~\cite{bdr}.  In order to seek similar solutions, we start with GSH parameters at GUT scale, 
but additionally provide $\mu$ and $m_A$ as inputs.  We run $m_{H_{u,d}}$ down, and at $Q = M_{SUSY}$ we 
compute what $m_{H_u},\,m_{H_d}$ should have been in order to give our input $\mu$ and $m_A$, and run 
back up using these new boundary conditions.  This weak scale Higgs (WSH) scenario has the parameters
\begin{equation}
m_{16},\,m_{10},\,M_D^2,\,m_{1/2},\,\tan\beta,\,m_A,\,\mu
\end{equation}

We take the GSH and WSH scenarios and search in their parameter spaces for regions having a good Yukawa 
unification where Yukawa unification is parametrized by $R = 
\frac{max(f_t,f_b,f_\tau)}{min(f_t,f_b,f_\tau)}$. 
Additionally we seek sub-regions that are consistent with the WMAP measurements of dark matter relic 
density $\Omega h^2$~\cite{bkss}.  There have been previous searches using the GSH input based on 
random scans and they were able to achieve less than few percent of Yukawa unification for 
$\mu>0$~\cite{abbbft}.  However the dark matter relic densities for these solutions were always much 
higher than the WAMP upper bound.  Here we implement the Markov Chain Monte Carlo (MCMC) technique which 
enables a much more efficient scanning of multi-dimensional parameter spaces.  The following two 
sections summarize the MCMC technique, the characteristics of the regions found by utilizing it, and 
dark matter-related consequences of the $SO(10)$ SUSY GUTs.

\section{The MCMC Search and the Yukawa-unified Solutions}

A Markov Chain is a discrete time, random process where given the present state, the future state only
depends on the present state, but not on the past states~\cite{mcmc}.  The MCMC samples from 
a given parameter space as follows: It takes a starting point, and it generates a candidate point $x^c$ 
from the starting point $x^t$ using a proposal density $Q(x^t;x^c)$.  The candidate point is accepted to 
be the next state $x^{t+1}$ if the ratio $p = \frac{P(x^c)Q(x^t;x^c)}{P(x^t)Q(x^c;x^t)}$ 
(where $P(x)$ is the probability calculated for the point $x$) is greater than a uniform random number 
$a = U(0,1)$.  If the candidate is not accepted, the present point $x^t$ is retained and a new candidate 
point is generated.  By repeating this procedure continuously the Markov Chain eventually converges at 
a target distribution around a point with the highest probability.  

Our MCMCs were directed to approach regions with $R \sim 1.0$ and $0.094 \le \Omega h^2 \le 0.136$.
Here we used a Gaussian distribution for the proposal density $Q$, and approximated the likelihood of a 
state to $e^{-\chi^2_{R,\Omega}}$.  We also chose multiple starting points $(\sim 10)$ in order to 
search a wider range of the parameter space.  We used {\tt ISAJET 7.75} for sparticle mass 
computations and {\tt micrOMEGAs 2.0.7} for DM relic density calculations.  The MCMCs successfully 
located some regions with good $R$ and $\Omega h^2$.  Here we will mostly emphasize the results from the 
GSH scenario.  

\begin{figure}
\begin{center}
\epsfig{figure=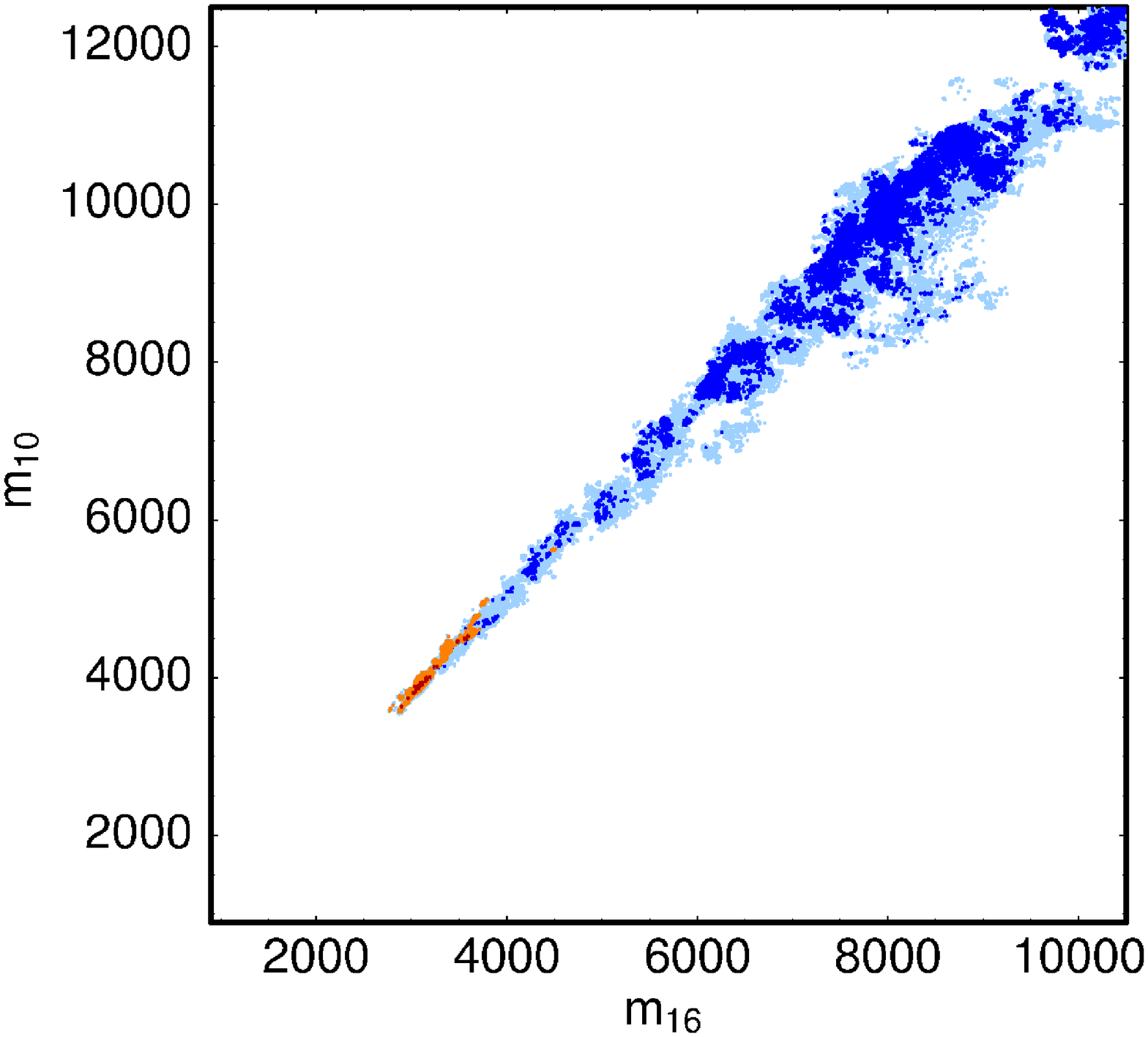,height=5cm} \qquad
\epsfig{figure=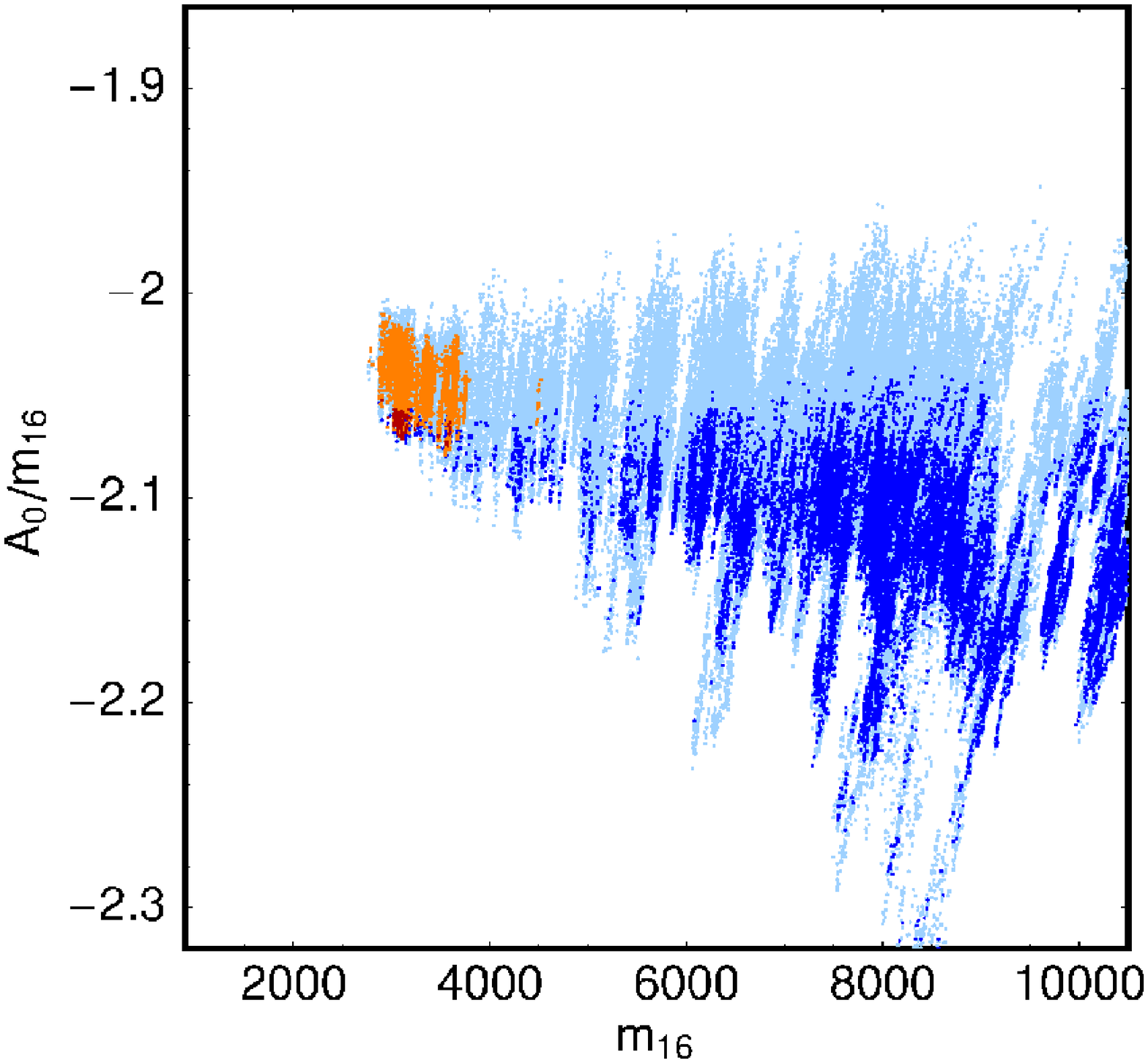,height=5cm}
\caption{Yukawa-unified GSH points found by MCMC in the $m_{10}$ vs. $m_{16}$
plane (left) and the $A_0/m_{10}$ vs. $m_{16}$ plane (right); the light-blue (dark-blue) points have 
$R<1.1\ (1.05)$, while the orange (red) points have $R<1.1\ (1.05)$ plus $\Omega h^2<0.136$.
\label{fig:GSHpar}}
\end{center}
\end{figure}

Figure~\ref{fig:GSHpar} shows the compatible regions in $m_{16}-m_{10}$ and $m_{16}-m_{16}/A_0$ planes.  
The light blue dots have $R \le 1.1$, the dark blue dots have $R\le1.05$, the orange dots have $R\le1.1$ 
plus $\Omega h^2 \le 0.136$ and the red dots have $R \le 1.05$ plus $\Omega h^2 \le 0.136$.  We see that 
Yukawa unification occurs only at the regions where the input parameters are strongly correlated, having 
$m_{10} \simeq 1.2m_{16}$ and $A_0 \simeq -(2 - 2.1)m_{16}$.  A good DM relic density is achieved only 
at the constrained regions that have $m_{16} \sim 3-4$~TeV.  Further search showed that $m_{1/2}$ takes 
the lowest possible values for a given $m_{16}$, generally giving $\sim 100$~GeV, and decreases steadily 
with increasing $m_{16}$. 

\begin{figure}
\begin{center}
\epsfig{figure=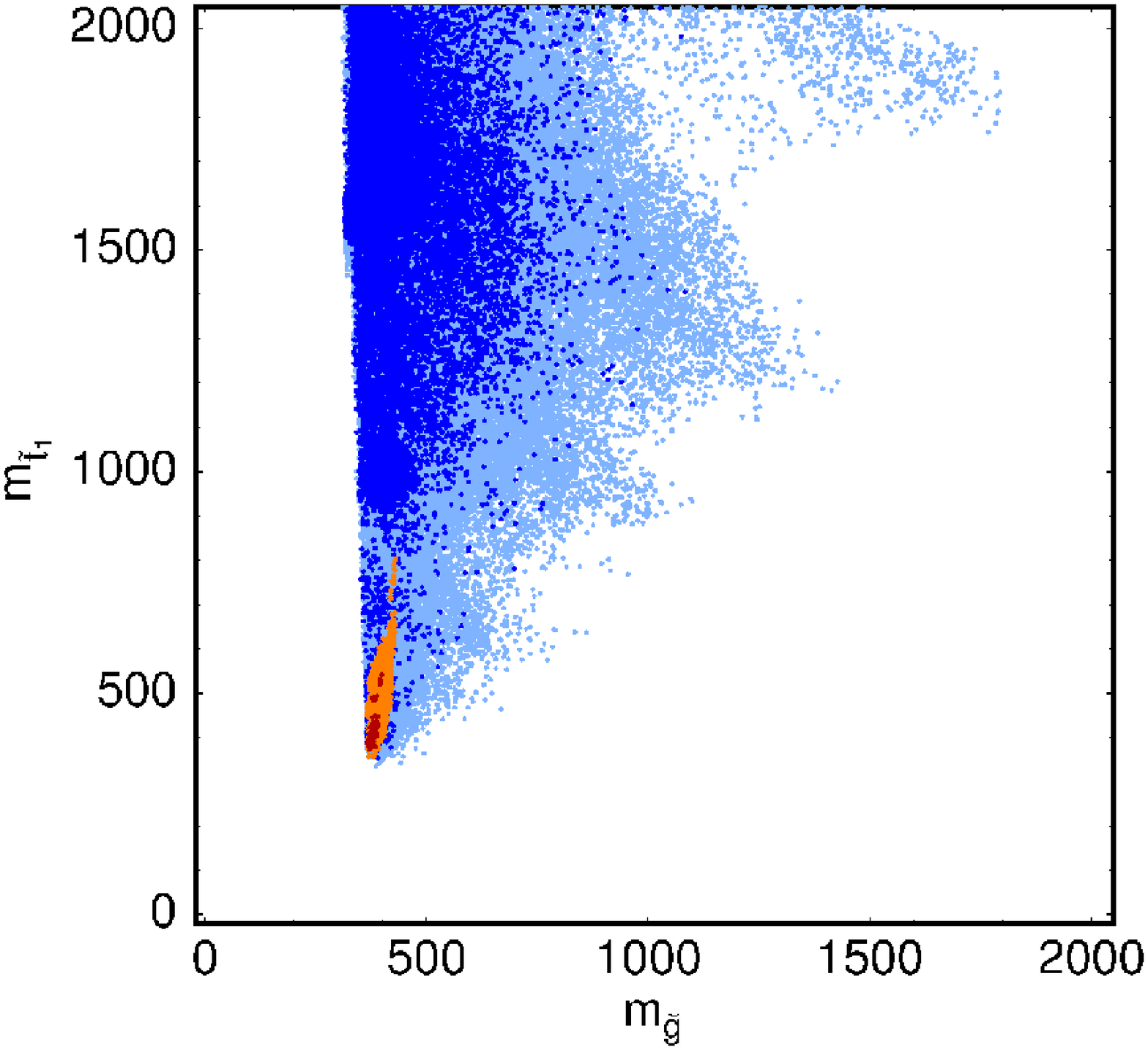,height=5cm} \qquad
\epsfig{figure=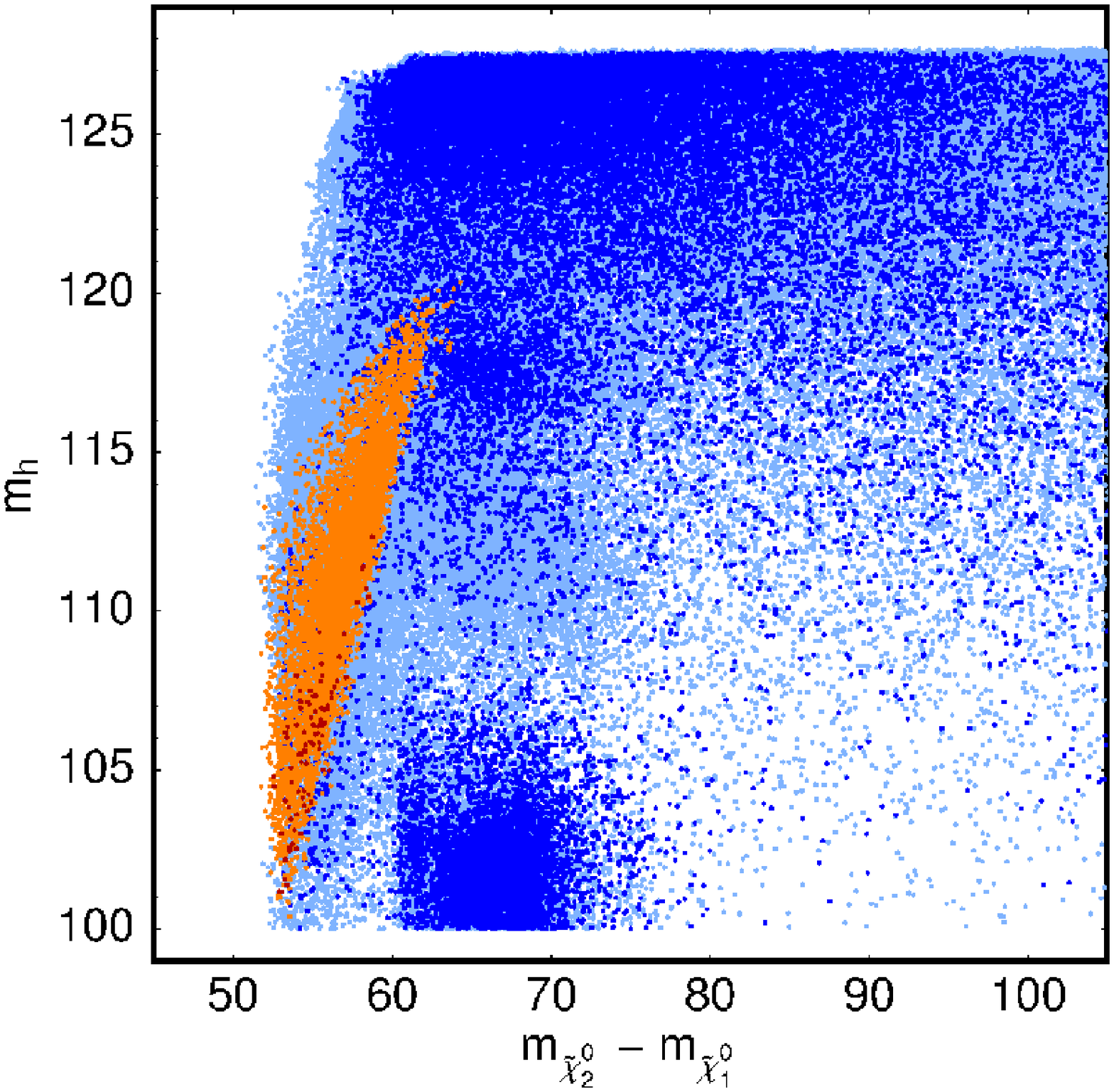,height=5cm}
\caption{Yukawa-unified GSH points found by MCMC in the $m_{\tilde{t}_1}$ vs $m_{\tilde{g}}$ plane 
(left) and $m_{h^0}$ vs. $m_{\tilde{\chi}^0_2} - m_{\tilde{\chi}^0_1}$ plane (right); color code as in 
Figure~\ref{fig:GSHpar}.
\label{fig:GSHobs}}
\end{center}
\end{figure}

These highly confined parameter regions lead to strongly constrained mass spectra, and hence to 
significant LHC signatures.  We see that Yukawa-unified solutions are distinguished by their 
heavy 1st/2nd generation scalars ($>2$~TeV), lighter 3rd generation scalars ($\sim$TeV) and light 
gauginos (few hundred GeV).  All Higgses except $h^0$ are about $1-3$~TeV.  Figure~\ref{fig:GSHobs} 
shows the distribution of selected points on $m_{\tilde{t}_1}$ vs $m_{\tilde{g}}$ plane (left), and 
on $m_{h^0}$ vs $m_{\tilde{\chi}^0_2} - m_{\tilde{\chi}^0_1}$ plane (right) for the GSH scenario.  
The requirement of $\Omega h^2 < 0.136$ favors a gluino mass range around $350-450$~GeV, which means we 
would expect a large amount of gluino pair production at the LHC with cross sections about $\sim 
100$~pb.  The gluinos decay via 3-body channels such as $\tilde{g} \rightarrow \tilde{\chi}^0_1 
b\bar{b}, \tilde{\chi}^0_2 b\bar{b}, \tilde{\chi}^\pm_1 t\bar{b}/b\bar{t}$, since 2-body channels are 
closed due to the high squark masses.  On the other hand favored $\tilde{\chi}^0_2 \simeq 
\tilde{\chi}^\pm_1$ mass range is $100-150$~GeV, which leads to gaugino pair production cross sections 
about $\sim 10$~pb, while $m_{\tilde{\chi}^0_1} \sim 50-75$~GeV.  The preferred mass difference 
$m_{\tilde{\chi}^0_2} - m_{\tilde{\chi}^0_1}$ is $52-65$~GeV which is smaller than $m_{Z,h^0}$, 
therefore $\tilde{\chi}^0_2$ decays are dominated again by 3-body channels as $\tilde{\chi}^0_2 
\rightarrow b\bar{b}\tilde{\chi}^0_1,q\bar{q}\tilde{\chi}^0_1,l\bar{l}\tilde{\chi}^0_1$.  
\setlength{\parskip}{0cm}

As a result we expect the $SO(10)$ models to manifest themselves as multi b-jet plus low missing 
$E_T$ final states at the LHC.  Additionally it would be possible to investigate the OS/SF dilepton 
channels where the dilepton invariant mass is bound by $m_{\tilde{\chi}^0_2} - m_{\tilde{\chi}^0_1}$ and 
try to reconstruct the $\tilde{g} \rightarrow b\bar{b} \tilde{\chi}^0_2 \rightarrow b\bar{b} l 
\bar{l}\tilde{\chi}^0_1$ cascades.

\section{Consequences for Dark Matter}

\begin{figure}
\begin{center}
\epsfig{figure=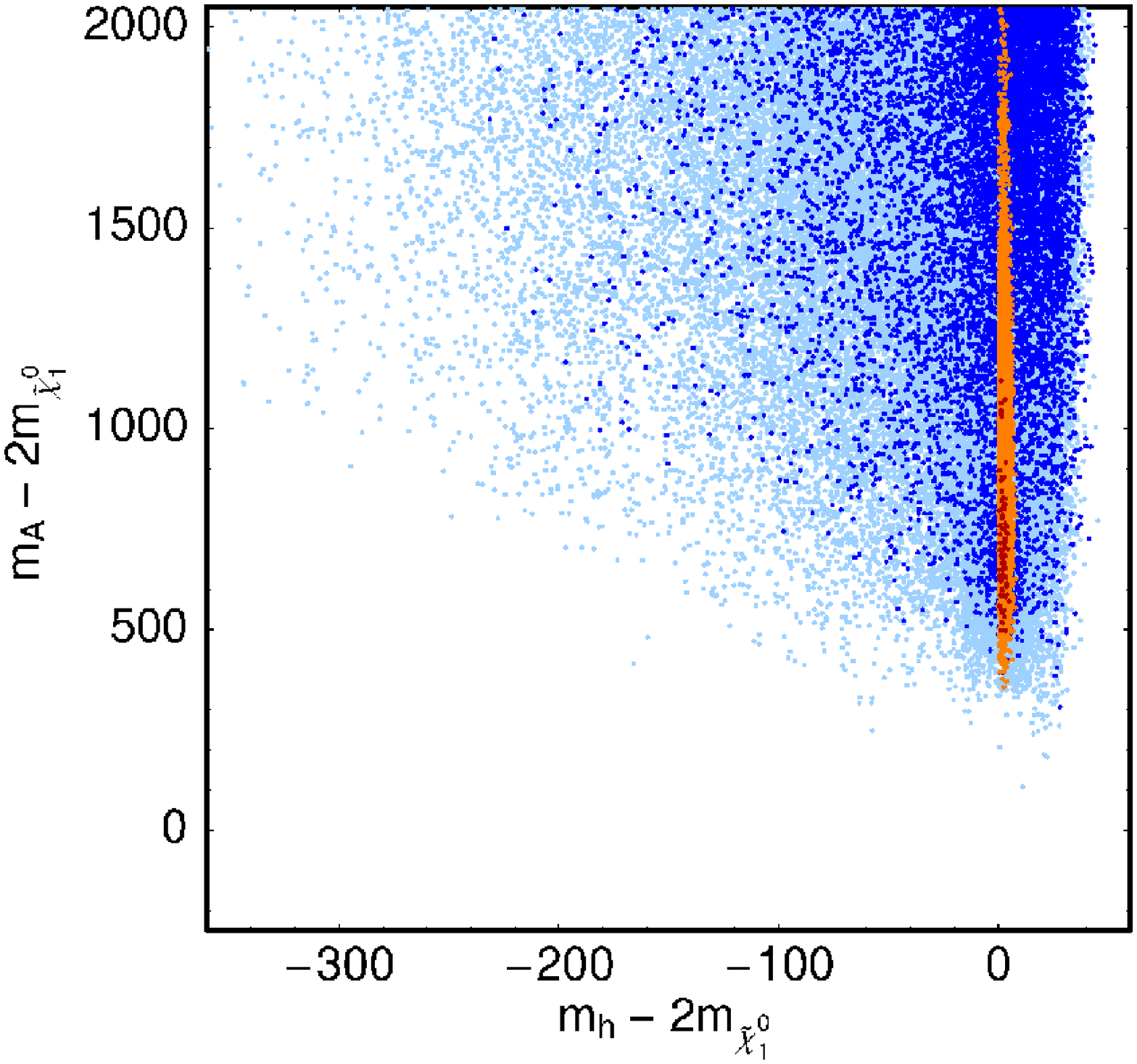,height=5cm} \qquad
\epsfig{figure=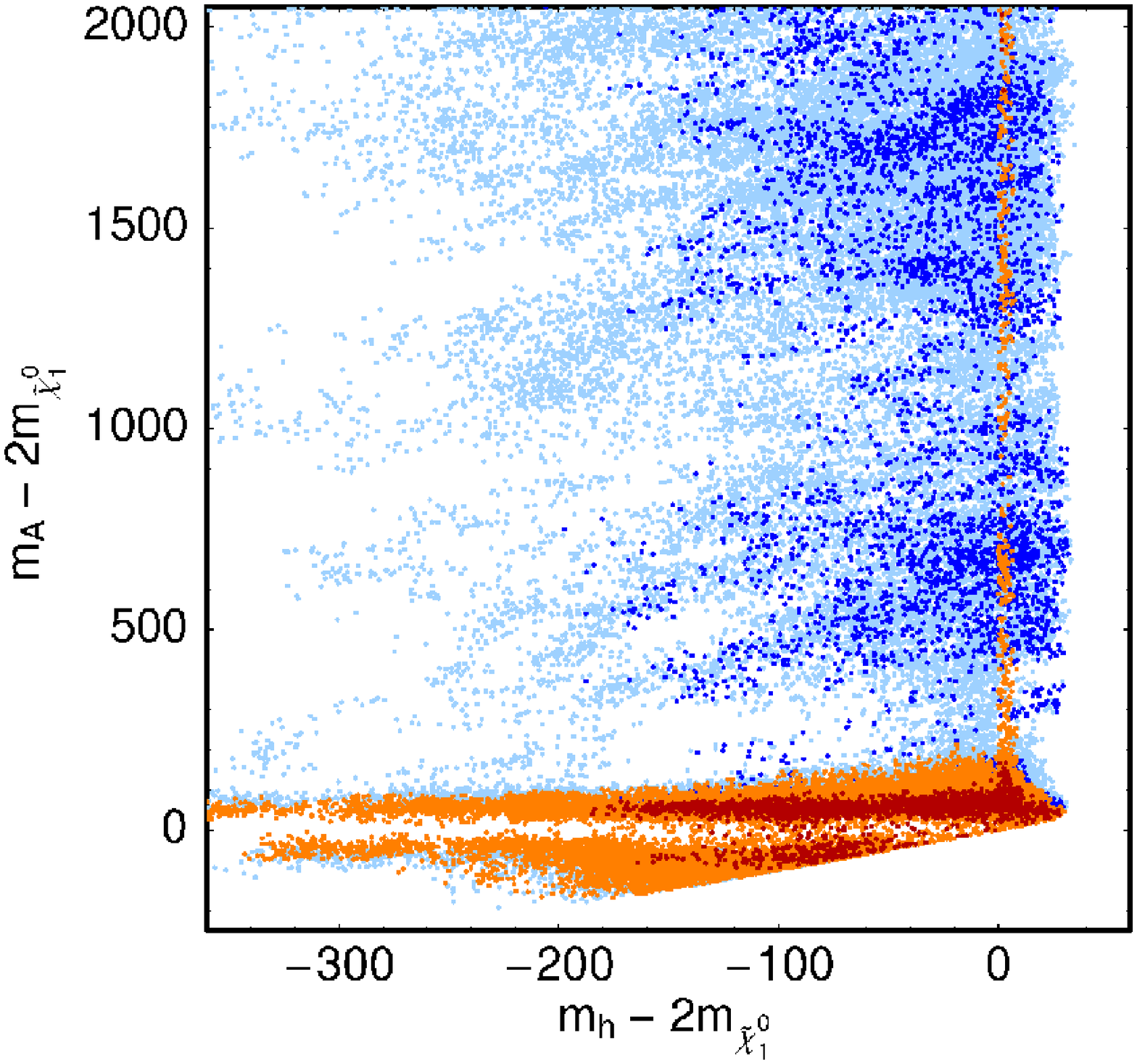,height=5cm}
\caption{Yukawa-unified points found by MCMC on the $m_A - 2m_{\tilde{\chi}^0_1}$ vs. $m_{h^0} - 
2m_{\tilde{\chi}^0_1}$ plane for GSH (left) nd WSH (right) scenarios; color code as in
Figure~\ref{fig:GSHpar}.  
\label{fig:dm}}
\end{center}
\end{figure}

The majority of solutions shown in Figures~\ref{fig:GSHpar} and~\ref{fig:GSHobs} have excess DM relic 
density, while only a small portion of them gives $\Omega h^2 < 0.136$.  To investigate the mechanism 
that provides the efficient annihilation in the DM-allowed solutions, we check the behavior of solutions 
in $m_{h^0}$ and $m_A$ mass resonances.  Figure~\ref{fig:dm} shows the distribution of Yukawa-unified 
points on the $m_A - 2m_{\tilde{\chi}^0_1}$ vs. $m_{h^0} - 2m_{\tilde{\chi}^0_1}$ plane for GSH (left) 
and WSH (right) scenarios.  In the GSH plot all DM-allowed solutions are on the $m_{h^0} \simeq 
2m_{\chi^0_1}$ line, which shows that the relic density is reduced by annihilation via a light Higgs 
resonance.  On the other hand $m_A > 2m_{\tilde{\chi}^0_1}$, so there are no A resonance solutions.  
Turning to the WSH scenario we see that annihilation via both $h^0$ and A resonances are at work.  
Actually the majority of solutions are generated by the latter due to the relatively small A masses 
allowed within the WSH scenario.  However all of these solutions have $B_s \rightarrow \mu \mu$ 
branching ratios higher than the latest reported CDF upper limit $5.8 \times 10^{-8}$.  So these A 
resonance points are ruled out, leaving us with only the $h^0$ resonance solutions.  

One could also devise alternative methods for reducing the excess DM relic density.  One way could be to 
assume that $\tilde{\chi}^0_1$ is not the LSP, but can decay to other candidates such as gravitino or 
axino via the mode $\tilde{\chi}^0_1 \rightarrow \gamma \tilde{G}/\tilde{a}$.  
Lifetime of $\tilde{\chi}^0_1$ would be long enough to let it escape the detectors.  The resulting 
relic density would be $\Omega_{\tilde{G},\tilde{a}} = 
({m_{\tilde{G},\tilde{a}}}/{m_{\tilde{\chi}^0_1}})\Omega_{\tilde{\chi}^0_1}$ 
since the $\tilde{G}s/\tilde{a}s$ inherit the thermally produced neutralino relic number density. 
$\tilde{G}$ LSP can only reduce the relic density a few times, which is not satisfactory for our case, 
but axinos with $m_{\tilde{a}} \le 1$~MeV would allow for a mixed cold/warm DM solution which can reduce 
the relic density below the WMAP bound~\cite{bs}.

Another method to reconcile $\Omega h^2$ is to relax some universalities in the GUT scale SSB terms.  
For example, increasing the $U(1)$ gaugino mass term $M_1$ (while keeping $M_{2,3}=m_{1/2}$) brings 
$m_{\tilde{\chi}^0_1}$ close to $m_{\tilde{\chi}^\pm_1}$, hence making $\tilde{\chi}^0_1$ more wino-like 
and inducing bino-wino coannihilation.  A further possibility is to lower the 1st/2nd generation masses 
$m_{16}(16)$ (while keeping $m_{16}(3)=m_{16}$), which enables neutralinos to annihilate via light 
$\tilde{q}_R$ exchange and leads to neutralino-squark coannihilation.  

\section{Conclusions}

By performing scans on the parameter space of simple $SO(10)$ SUSY GUT scenarios using the MCMC 
technique, we showed that solutions with both $5-10\%$ Yukawa unification and WMAP-compatible $\Omega 
h^2$ can exist around $m_{16}\sim 3-4$~TeV.  These regions defined by strictly constrained relations 
among the GUT scale inputs generate special sparticle mass relations that lead to distinguishable 
signatures at the LHC.  With multi-TeV scalars, $350-450$~GeV gluinos and $50-150$~GeV light gauginos, 
we expect dominant gluino production followed by 3-body cascade decays which will end up in b-rich 
multijet final states, occasionally including OS/SF lepton pairs from $\tilde{\chi}^0_2$ decays.  
Moreover the possibility to lower $\Omega h^2$ by assuming $\tilde{a}$ LSP or introducing SSB 
non-universalities marks wider parameter space regions as compatible.  So we can conclude that $SO(10)$ 
SUSY GUTs provide motivated scenarios with robust signatures relevant to be tested soon at the turn-on 
of the LHC.  

\section*{Acknowledgments}

I thank the organizers of XLIIIrd Rencontres de Moriond for a fulfilling conference and gratefully 
acknowledge the financial support by the EU Marie Curie Actions project.

\end{document}